\documentclass[,numbers]{aipproc}

\layoutstyle{6s}

\usepackage{graphicx}
\usepackage{amsmath}
\usepackage{subfigure}
\usepackage{sidecap}

\title{A Search for Short-term Variability in the Very High Energy $\gamma$-ray Emission from the Crab Nebula}
\author{Anna O'Faol\'ain de Bhr\'oithe$^*$ for the VERITAS Collaboration}{address={$^*$School of Physics, University College Dublin, Belfield, Dublin 4, Ireland},email={anna.o-faolain-de-bhroithe@ucdconnect.ie}}

\begin{abstract}
The Crab Nebula has long been considered a standard candle in high energy astrophysics, but in recent years
 this assumption has been strongly contradicted in keV-GeV wavebands. In light of these developments, a
 search for variability is being performed on the nebula at Very High Energies (VHE; $E>300$ GeV), the
 preliminary results of which are presented here. This initial study is based on 10 years ($2001-2011$) of
 archival data from the Whipple 10\,m telescope. The data set was searched for evidence of variability on the
 timescales of 1, 7, and 14 days. To date, no significant flaring activity has been found, but simulations are
 in progress to determine the level of variability that would be detected.
\end{abstract}

\begin{document}
\maketitle

\section{Introduction}
The Whipple 10\,m $\gamma$-ray telescope is located at the Fred Lawrence Whipple Observatory in southern Arizona.
 Until June 2011, it operated in the range $0.3 - 10$ TeV and pioneered the Imaging Atmospheric \v{C}erenkov
 Telescope (IACT) technique for the detection of VHE $\gamma$-rays. The telescope was of Davies-Cotton design,
 consisting of a reflector and a camera at the focal plane to record the $\gamma$-ray images. The reflector was
 composed of 248 tessellated hexagonal mirrors mounted on a spherical surface with a total reflecting area of
 $\sim75$ m$^2$. The last camera in operation consisted of 379 PMTs and had a field of view of $\sim2.6^\circ$
 with an angular resolution of $0.117^\circ$, as described in~\cite{whipple_specs}.

Detected by the Whipple 10\,m telescope, the Crab Nebula was the first TeV source discovered~\citep{whipple_crab}.
 It has since been considered a standard candle of VHE $\gamma$-ray astronomy, as it has for X-ray and lower
 energy $\gamma$-ray astronomy~\citep[e.g.,][]{x-ray_crab}. As the brightest source in the VHE sky, it was ideal
 for this role.

In 2011, both AGILE~\citep{agile} and \emph{Fermi}-LAT~\citep{fermi_flare} reported the discovery of flaring
 activity in the Crab Nebula at MeV - GeV energies. The flares occur at a frequency of $\sim1-2$ per year, and
 have been observed to last between 4 and 15 days. ARGO-YBJ~\citep{argo} also reported enhanced flux at GeV -
 TeV energies from the Crab Nebula. These developments have motivated this search for VHE variability on short
 timescales for which the Atmospheric \v{C}erenkov Technique is particularly suited.

\section{Analysis and results}
A data set of Crab Nebula observations taken with the Whipple 10\,m telescope was compiled from the last 10
 years ($2001 - 2011$). Data were taken in 28-minute observations with a standard experimental setup under
 good weather conditions with an elevation angle $>55^\circ$.

Motivated by the flaring timescales reported by \emph{Fermi}-LAT and AGILE, the Whipple data set was searched
 for short-term variability on timescales of 7 and 14 days, as well as a shorter timescale of 1 day. A sliding
 window algorithm was developed to perform this analysis. The window is shifted along the data set night-by-night
 for each season. The significance of the emission in the window is calculated using the search window as the
 \texttt{ON} region and the the rest of the season as the \texttt{OFF} region. A high significance value would
 indicate the presence of a flare. 

\begin{SCfigure}
\centering
\includegraphics{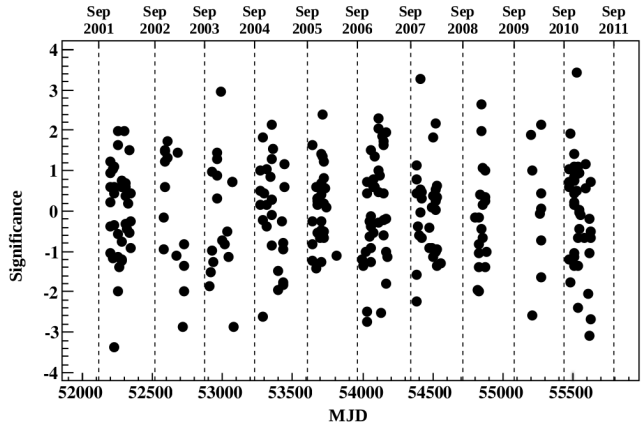}
\caption{1-day search window significances for the 10 years of Whipple data analysed. No significant periods of
 elevated emission are present in this data set.}
\label{sigs}
\end{SCfigure}

Figure~\ref{sigs} shows the window significances for a search timescale of 1 day for the 10 observing seasons.
 The highest significance detected in a 1-day window was $3.42\sigma$ pre-trials, corresponding to a post-trials
 probability of $0.078$ ($\sim2.07\sigma$)~\citep[e.g.,][]{trials}. Both 7- and 14-day search windows yielded
 lower post-trials maximum significances.
% of $1.38\sigma$ and $1.59\sigma$ respectively.
 Thus, there is no evidence for strong VHE flaring activity on these timescales in the current data set.

\begin{SCfigure}
\centering
\includegraphics{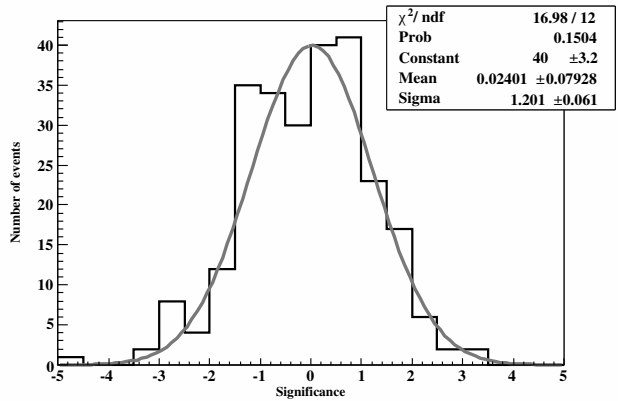}
\caption{Distribution of 1-day window significances, obtained by histogramming the data shown in Figure~\ref{sigs}
 and fitting with a Gaussian.}
\label{datasigdist}
\end{SCfigure}

Figure~\ref{datasigdist} shows the distribution of window significances obtained with a 1-day search timescale
 for the 10-year data set. A similar distribution was created for each search window timescale, and fit with a
 Gaussian. The variances of the Gaussian fits to the significance histograms are not consistent with 1.0,
 indicating that the observed variations are not solely due to statistical fluctuations. Randomising the dates
 of the observations and reanalysing the ``shuffled'' data preserves the width of the distribution, so it is
 independent of the configuration of the data.
% The non-uniform zenith angle of observations may contribute to
% this phenomenon. Atmospheric changes are another likely cause, but cannot be corrected for due to the lack of
% rigorous on-site atmospheric monitoring.

\section{Simulations}
A Monte Carlo simulation was developed to test whether the width of the significance histograms are consistent
 with statistical fluctuations, given the overlapping search windows. 18,000 nights of observations
 were simulated with the same source sampling as the real data and analysed. It was found that
 significance histograms produced with the three search timescales from the simulated data have variances very
 close to 1.0.

25,000 individual data sets, equivalent in length and sampling to the observational data, were then simulated and
 analysed and a distribution of the variances was produced. A variance of 1.3 (as seen in the real data, see
 Figure~\ref{datasigdist}) was observed in only two cases. This clearly points to a non-statistical source of the
 broadness of the data distributions. Varying observation angles and atmospheric changes are likely the main
 contributing factors.

The simulation was adapted to simulate a single flare of known length and emission within an otherwise standard
 data set. The data sampling was adjusted to ensure one simulated run per night for the duration of the flare,
 while still maintaining random sampling in the rest of the data set. This idealised scenario of full sampling of
 the flare provides the means to put an upper limit on the level of flaring activity that would be detected. The
 simulation was run 600 times for two different flare emission levels. In both cases, a medium flare duration
 of 5 days was used, with flare emission levels of 2 $\times$ average Crab Nebula and 1.5 $\times$ average Crab
 Nebula.

\begin{figure}
\centering
\includegraphics{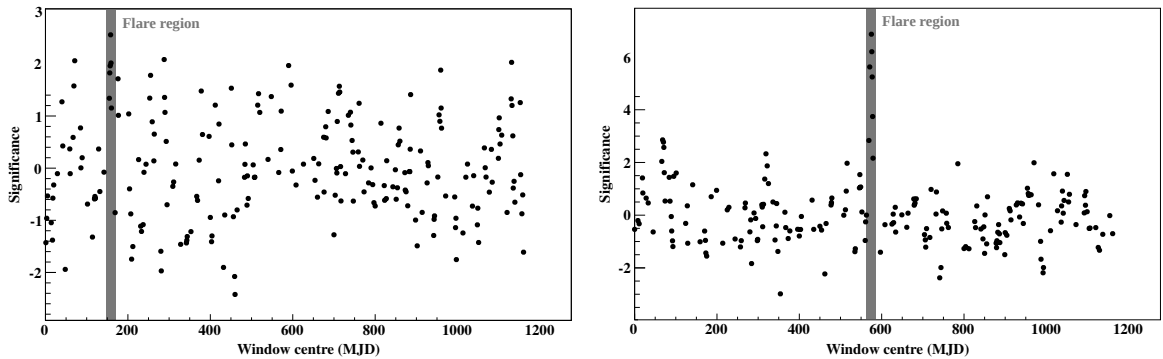}
\caption{Window significances for simulated flare of 5-day duration: the left panel shows a flare with a
 $50\%$ increase in emission over average levels and the right panel shows a flare with a $100\%$ increase
 in emission over average levels. While the flare in the left panel cannot easily be seen, the flare in
 the right panel is clearly visible.}
\label{flares}
\end{figure}

Figure~\ref{flares} shows typical data sets obtained for both flare emission levels. For a
 7-day window, it was found that the $\times2$ flare was detected above $5\sigma$ post-trials in $69\%$ of
 the data sets, and even when not detected it was always clearly visible by eye. The $\times1.5$ flare was only
 detected once post-trials, and was generally impossible to pick out by eye. 

\section{Conclusion}
No significant flaring activity has been found in this 10-year archival data set from the Whipple 10\,m telescope.
 The recent model of Bednarek and Idec~\citep{bednarek} predicts TeV flux variability of the order of $\sim10\%$
 above 1 TeV and on the same timescales as that observed at GeV energies. The Monte Carlo simulations indicate
 that flares would need to be of the order of $\sim100\%$ in order to be significantly
 detected, and so this model cannot be constrained with the current data set. However, this work will be expanded by
 extending the Whipple archival data set to include earlier epochs, which could potentially double the data
 set. VERITAS data will also be added to the study, considerably augmenting the data set from 2007 onwards and
 potentially providing the sensitivity to constrain the emission model.

 A search for a long-term decline in the TeV Crab Nebula flux, similar to
 that seen at keV energies~\citep{crab_decline} will be undertaken. This is complicated by the fact that the Crab
 Nebula has been used as a calibration source for IACTs since the founding of this field.

\section{Acknowledgements}
\small
This research is supported by grants from the U.S. Department of Energy Office of Science, the U.S. National
 Science Foundation and the Smithsonian Institution, by NSERC in Canada, by Science Foundation Ireland
 (SFI 10/RFP/AST2748) and by STFC in the U.K. We acknowledge the excellent work of the technical support staff
 at the Fred Lawrence Whipple Observatory and at the collaborating institutions in the construction and operation
 of the instrument. Anna O'Faol\'ain de Bhr\'oithe acknowledges the support of the Irish Research Council
 ``Embark Initiative''.

\end{document}